\documentclass[twocolumn,showpacs,preprintnumbers,amsmath,amssymb]{revtex4}
\usepackage [dvips]{graphicx}

\begin{document}

\title{Simple Model for Energy and Force Characteristics of Metallic
Nanocontacts\footnote{Ukr. Fiz. Jour. 2003 in print}}
\author{V.V. Pogosov$^{1,2}$\footnote{Corresponding author: E-mail:
 vpogosov@zstu.edu.ua (V.V.~Po\-go\-sov)},
V.P. Kurbatsky$^1$, D.P. Kotlyarov$^1$, A. Kiejna$^2$}
\address{$^1$Department of Microelectronics, Zaporozhye National
Technical University, Zaporozhye 69064, Ukraine\\
$^2$Institute of Experimental Physics, University of Wroc{\l}aw,
Wroc{\l}aw 50-204, Poland}

\date{\today}

\begin{abstract}
The quantum size oscillations of the energetic properties and the elongation
force of the gold slabs and wires, isolated and in a contact with
electrodes, are calculated in a free-electron model. A simple relation
between the Fermi energy and the square-potential-well depth is used and
tested for low-dimensional systems. It is shown that considering the
electron subsystem of a slab (or wire) in a contact as open one, the contact
acts like a sort of electron pump which sucks or pumps out electrons from
the sample. The effect of the contact potential difference on the elastic
force oscillations is considered. The calculated amplitudes of force
oscillations are in a qualitative agreement with those observed
experimentally.
\end{abstract}

\maketitle

\section{Introduction}

Recent experimental investigations of gold point-contacts by using
the atomic force microscope \cite{1} have demonstrated a violation
of the Hooke and Ohm laws: a deformation of the contact leads to a
coherent stepwise variation of the conductance and force.
Theoretical interpretation of these observations for a deformed
nanowire was given  in the framework of the simple free-electron
models \cite{2,3,4,5,6,6a}. The energetic picture of the
nanocontact is more complicated, however, than that adopted in
\cite{2,3,4,5,6}. First, in the previous considerations the
contact potential difference \cite{7} is not taken into account.
Secondly, the metallic contact changes its dimensionality with the
variation of sample-cantilever distance. At first, its shape is
rather similar to a slab, to resemble a wire at the moment of
rupture \cite{8,9}. Thus, we have to do with a transformation of
the 2D open electron systems into the 1D one. The purpose of this
work is to investigate the limiting cases of this transition. We
study the size dependence of the work function, ionization
potential and elastic force of the isolated quantum slab and wire
that are inserted  between two electrodes to form a contact. We
begin with consideration of the energetics of an isolated sample.
The results are exploited later to determine the contact potential
difference (CPD) for a sample connected with electrodes.

In order to trace analytically the effect of dimensionality on the
energetic and force properties one needs a simple model relating
mutually Fermi energy, work function and the potential barrier.
We consider a rectangular sample of the volume $V=a\times a\times L$, where $%
L$ denotes the dimension along the $x$-axis. The inequalities $a\gg L$ and $%
a\ll L$ correspond to the geometry of a slab and a wire, respectively. The
consideration of these two asymptotic limits which simulate the early and
late phases of elongation, allows us to trace the evolution of the energy
and force characteristics of the 2D and 1D metallic structures.

We have assumed that the energetics of a finite system of bounded
electrons can be described by a square potential well of the width
$L$ (slab) or $a$ (wire) and the depth $U<0$,
\begin{equation}
-U(E_{F})=W(E_{F})+E_{F},  \label{U}
\end{equation}
where $W$ is the work function and $E_{F}$ is the Fermi energy of
a finite sample.

Let us remember that the potential well of the depth $U$ of metal
sphere shows the size oscillations \cite{10}. The Fermi energy of
metal film also has a size oscillations \cite{11}. Due to an
expansion of the chemical potential of spherical cluster of $R$
radius $\mu =\mu_0 + {\mu_1}/R$ in powers of the inverse radius
\cite{12} we paid attention on the inequality
\begin{equation}
        W = W_0 - \frac{\mu_1}{R}  < W_0 .           \label{WN2}
\end{equation}

The criterion $W<W_{0}$ (subscript zero labels the quantities for
a semi-infinite metal) for one-dimensional systems was introduced
by us \cite{13} to check the correct size dependence of the
$U(E_{F})$ (by means $E_{F}(L)$). In order to determine the size
oscillations in the potential well depth $U(E_{F})$ we have
employed a simple expression for the $W_{0}(E_{F0})$ provided by
\cite{14}. The latter is based on the concept of the image force
action \cite{15,16} and a spontaneous polarization of metallic
plasma, to determine the distance, from the classical metal
surface at which the image force begins to act (see also
\cite{16a} and references therein),
\begin{equation}
W_{0}(E_{F0})=\frac{B}{r_{s}^{3/2}E_{F0}^{1/2}}, \quad
\mbox{(a.u)} \label{WEF1}
\end{equation}
where $B$ is the adjusting parameter and $r_{s0}$ is the average
distance between electrons in the bulk, $B=0.3721$ a.u. and
$r_{s}=$3.01 bohr radius for gold.

Adapting the result for the low-dimensional systems we replace
$W_{0}(E_{F0})$ by $W(E_{F})$ and suppose  $W\rightarrow W_{0}=$
4.30 eV \cite{39} as $L\rightarrow \infty$ \cite{13}. In order to
get a qualitative agreement of calculated force with the
experimental data we have to assume that Au is monovalent
\cite{2,3,6}.

\section{Basic Relations for an isolated sample}

For the 2D and 1D metallic systems the allowed energy levels (the
electron kinetic energies) form a quasi-continuum \cite{40},
$$
E_{n_{xyz}}=E_{n_{x}}+E_{n_{y}}+E_{n_{z}}=\frac{\hbar
^{2}}{2m}(k_{n_{x}}^{2} +k_{n_{y}}^{2} +k_{n_{z}}^{2}).
$$
For an infinitely deep potential well of the width $a$ (the shape
is a cube) this expression reduces to $E_{n_{xyz}}= \hbar^2 \pi^2
n_{xyz}^2/(2ma^2)$ with $n_{xyz}^{2} =
n_{x}^{2}+n_{y}^{2}+n_{z}^{2}$.

For simplicity, we assume that the wave vector components  are
the solutions of the transcendental equations for square potential well:
\begin{equation}
    k_{n}M =n\pi -2\arcsin (k_{n}/k^{0}),  \label{arc}
\end{equation}
where $n = 1, 2, 3,\ldots$, and $n=n_{x}$ for $M=L$ and  $n=n_{y},n_{z}$ for $M=a$,
$\hbar k^{0}=\sqrt{-2mU}$. One gets two relations of identical form, to determine
$k_{n_{y}}$ and $k_{n_{z}}$. The inequalities $\max \{n_{y},n_{z}\}\gg \max \{n_{x}\}$
for slab and $\max \{n_{y},n_{z}\}\ll \max \{n_{x}\}$ for wire correspond
to highest occupied level.

The density of electron states, $D(E)$, is defined by the sum
$\sum _{n_{xyz}} \delta(E-E_{n_{xyz}})$ over the allowed states.
Replacing the three-dimensional summation in the $k$-space by
integration over $k_y$ and $k_z$ (or over $k_x$) and summation
over $n_{x}$ (or over $n_{y}$ and $n_{z}$), we get
\begin{equation}
  D(E)=\frac{m}{\pi L \hbar^2}n_{E},                     \label{DOS1}
\end{equation}
and
\begin{equation}
  D(E)=\frac{L}{V}\sqrt{\frac{2m}{\pi^2\hbar^2}}\sum_{n_{y},n_{z}}^+
                 (E-E_{n_{y}}-E_{n_{z}})^{-1/2}                     \label{DOS}
\end{equation}
for the finite slab (or wire, respectively).

In the equation (\ref{DOS1}) for a slab $n_{E}$ is the integer part of number,
\begin{equation}
n_{E}=\left[ \frac{kL+2\arcsin (k/k^{0})}{\pi }\right] , \label{+}
\end{equation}
where $\hbar k=\sqrt{2mE}$.

In the equation (\ref{DOS}) for a wire the plus sign in the limit of summation
indicates that $n_{y}$ and $n_{z}$ run
from 1 to the maximum value for which the expression under the square root is
positive.

Subsequently, the total number of electrons in a slab is given by
\begin{equation}
N=\frac{a^{2}m}{\pi \hbar^2}\sum\limits_{n_{x}=1}^{n_{F}}(E_{F}-E_{n_{x}}),
  \label{N}
\end{equation}
where $n_{F}$ is the number of highest occupied subband, $n_{F}$ equals $n_{E}$
 in Eq.(\ref{+}) with the change $k\rightarrow k_{F}$.

The total number of electrons in a wire is given by
\begin{equation}
   N= 2L \sqrt{\frac{2m}{\pi^2\hbar^2}}\sum_{n_{y},n_{z}}^+
                         (E_F-E_{n_{y}}-E_{n_{z}})^{1/2}   .         \label{N2}
\end{equation}
Here $N= \overline{n}V$, where $\overline{n}=3/4\pi r_{s0}^{3}$ is
the electron density in the bulk of semi-infinite metal.

Discussing the effect of dimensionality  and charging, it is
useful to analyze ionization potential which is a well defined
quantity for an arbitrary size of a sample \cite{30,30a}. The
ionization potential which is defined as a work needed to remove
an electron from a neutral metallic sample, can be expressed as
\begin{equation}
   IP=  W + \frac{e^2}{2C},  \label{IP}
\end{equation}
where $C$ is the capacitance of the sample. It should be mentioned
that $W_0$ and $IP$ are the experimentally measured quantities,
while $W$ has mainly a methodical sense because it can be measured
only in the limit of $C\rightarrow \infty$. An extended thin slab
or wire of infinite length has an infinite capacitance,
$C\rightarrow\infty$, therefore $IP \rightarrow W$. Equation
(\ref{IP}) can be interpreted as the effect of charging on the
work function of the neutral finite sample. It should be noted
that the size correction in $W(L)$ which is similar to that of the
spherical cluster (see Eq.(\ref{WN2})) competes with the term
$e^2/2C$. Since it is impossible to derive an analytical
expression for the rectangular sample, in order to estimate the
monotonic size dependence of $IP$ we apply the well-known formula
for the capacitance $C$ of the disk of the width $L$ and of the
needle of the length $L$ \cite{18}.

Solving the set of Eqs.(\ref{arc}) and (\ref{N}) or (\ref{N2}), by
using (\ref{U}) and adapted Eq.(\ref{WEF1}), one can determine the
Fermi energy $E_{F}$ and subband energies for a given $L$ under
condition $V=$ constant. Consequently, one gets the work function
$W$, as function of the Fermi energy of finite sample.

The elastic elongation force acting on a finite sample is given by
$F=-dE_{t}/dL$, where $E_{t}$ is the total energy of the sample.
Neglecting the temperature effects in the adiabatic approximation,
from the virial theorem we have $E_{t}=-K$, which means that the
energy of a bound electron-ion system is negative \cite{13,19}.
(The virial and stress theorems for electron--ion mixtion of solid
metal representation was formulated in Ref. \cite{Z}).
Consequently,
\begin{equation}
F=\frac{dK}{dL}.  \label{F1}
\end{equation}
In expression (\ref{F1}) we take into account the potential energy
of the electron-ion system. This gives a plus sign in front of $K$
(an opposite sign appears in \cite{2,3,6}, where the simplest
model of nanowire has been exploited). The plus sign follows from
the application of the virial theorem.

The total kinetic energy of electrons are given by the expressions
\begin{multline}
K=\frac{a^{2}}{2\pi ^{2}}\frac{\hbar ^{2}}{2m}\sum_{k_{n_{x}}}\int%
\limits_{0}^{\sqrt{k_{F}^{2}-k_{n_{x}}^{2}}}(\kappa ^{2}+k_{n_{x}}^{2})2\pi
\kappa d\kappa =\\
\frac{a^{2}m}{2\hbar ^{2}}\sum%
\limits_{n_{x}=1}^{n_{F}} (E_{F}^{2}-E_{n_{x}}^{2}),  \label{KE}
\end{multline}
and
\begin{multline}
 K = \frac{2L}{3} \sqrt{\frac{2m}{\pi^2\hbar ^2}}
  \sum_{n_{y},n_{z}}^+
(E_F - E_{n_{y}} - E_{n_{z}})^{1/2}\times\\
(E_F + 2E_{n_{y}} +
2E_{n_{z}}),  \label{K}
\end{multline}
for an isolated slab and wire, respectively.

\section{The effect of a \lq\lq\,point\,\rq\rq contact}

When a sample is inserted in a contact between the reservoirs then
its electron subsystem has to be considered as an open one, under
the condition $W(L)=W_0$, where $W$ and $W_0$ are the work
functions for an isolated sample and reservoirs, respectively. Due
to the contact potential difference, $\delta \phi$, the
electroneutrality of the sample breaks down and $\delta N$
electrons from the electron subsystem are transferred into
reservoirs. In order to determined $\delta \phi $ one can imagine
a simple energetic cycle in which electronic charges are
transferred from a sample to infinity and then into electrodes. By
expressing the ionization potential, $IP$, of the sample charged
by $+e\delta N$ in a form similar to that for the charged
spherical metal clusters \cite{20,21} we have,
\begin{multline}
IP=E_{N-\delta N-\Delta }-E_{N-\delta N}=\\
W\Delta+\frac{e^{2}}{2C}[(\delta N+\Delta )^{2}-\delta N^{2}],
\label{PC}
\end{multline}
where $-e\Delta$ is the part of the electronic charge which leaves
the charged sample. The electron affinity of this electronic
charge, transferred to reservoirs is $EA=W_{0}\Delta$. The
equilibrium condition $IP-EA=0$ leads to
\begin{equation}
W_{0}-W-\frac{e^{2}}{2C}(2\delta N+\Delta )=0.  \label{PC2}
\end{equation}
It should be noted that the magnitude of $\Delta $ can be
infinitesimally small (non-integer), because in a contact (open
system), the residual electron can be transferred only partially
(i.e., there is a finite probability that it could be found both
in a sample and in a reservoir). Thus, we can ascribe to $\delta
N$ a continuous value. We also suppose that $C$ of rectangular
sample appearing in Eq.(\ref{PC}) corresponds to the total
capacitance $C_{c}$ of both contacts. This is justified by the
fact that near the two faces of a sample the surplus positive
charges have a similar surface distribution both in the case of
real ionization of an isolated sample and for a sample in contact.
(This situation is similar to that encountered in single-electron
devices \cite{22}). Then, assuming that $C_{c}=e\delta N/\delta
\phi $, $\delta N\ll N$ and $\Delta \rightarrow 0$, Eq.(\ref{PC2})
one gets
\begin{equation}
\delta \phi =(W_{0}-W)/e>0.  \label{PC3}
\end{equation}
For example, the energy spectrum $E_{n1_{x}}$ of the remaining electrons
in the slab, $N_{1}=N-\delta
N$, can be found by solving (\ref{arc}) for a square potential well of depth
\begin{equation}
U_{1}=U-e\delta \phi ,  \label{PCF}
\end{equation}
where $U$ corresponds to the isolated slab. Comparing (\ref{PCF}) with (%
\ref{U}) and using (\ref{PC3}) we find that in equilibrium between a sample
and the electrodes, the magnitude of $E_{F1}$ is equal to that of an
isolated sample ($E_{F}$). The value of $\delta N$ can be calculated using
Eq.(\ref{N}) with the change $N\rightarrow N_{1}$ and $k_{n_{x}}\rightarrow
k_{n1_{x}}$. The total kinetic energy, $K_{1}$, of the remaining electrons
is defined as earlier (Eq.(\ref{KE})) with the changed energy spectrum and
the number of electrons. The same approach was used for the nanowire.
In the case of an open system the elastic force is
determined by a surplus pressure, relative to the reservoirs, multiplied by
the area of the contact:
\begin{equation}
F_{1}=-\frac{d\Omega }{dL},  \label{PCF1}
\end{equation}
where the size-dependent part of the grand potential $\Omega $ is given by
\begin{equation}
\Omega =E_{1}+W_{0}N_{1},  \label{PCF2}
\end{equation}
and $E_{1}=-K_{1}$.


\section{Results and discussion}

The calculations were performed for the set of isolated Au slabs
and wires and then for the ones in contact with reservoirs. It
allowed us to calculate the contact potential which is needed for
the force characteristics. Assuming the ideal plastic deformation
in the experiments \cite{1,9}, the volume of deformed sample will
be constant. All considered samples have a same volume, $V=4$
nm$^{3}$, and the number of
electrons $N=236$. The linear sizes of a sample vary in the range: $\sqrt{%
\pi }r_{s}<L<13a_{0}$ for slabs, and $L_{0}/10>a>\sqrt{\pi}r_{s}$
for wires. Here, $a_{0}=\hbar ^{2}/me^{2}$ is the Bohr radius. The
calculated energy and force characteristics corresponding to a
slab and wire, respectively, are displayed in the left and right
parts of Figs.1 and 2. The wire length, $L=L_{0}+\Delta L$, has
been increased by about seven times.
\begin{figure}[!ht]
\centering
\includegraphics [width=0.45\textwidth] {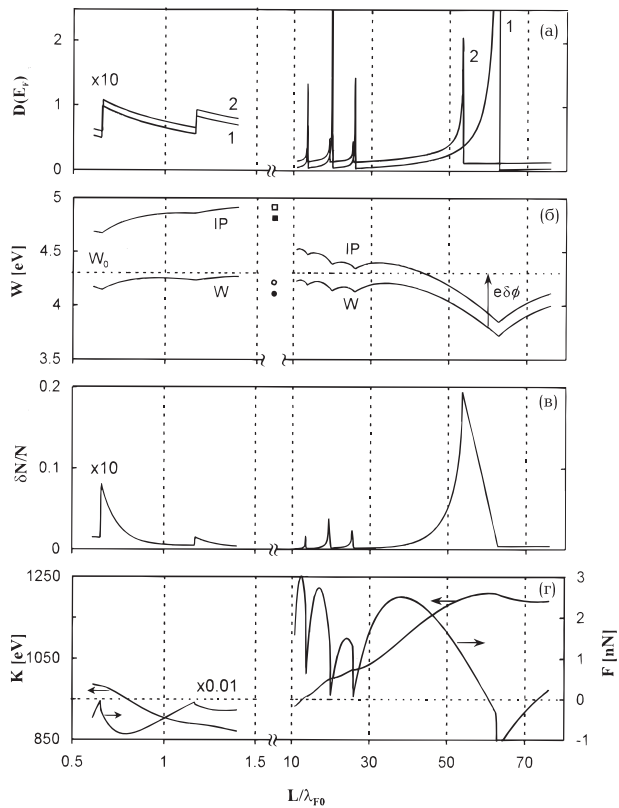}
\caption{Characteristics of the elongated sample as functions of
the length (in units of $L/\lambda _{F0}$, where $\lambda _{F0}$
is the Fermi wave length of the reservoir, $\lambda
_{F0}=2\pi/k_{F0}$). The left side of the figure corresponds to a
slab while the right one to the wire. (a) Density of states at the
Fermi level of a isolated sample (curves 1) and sample in a
contact (curves 2). (b) The work function, ionization potential,
and contact potential difference of the isolated slab and wire.
For comparison, the values of $W$ ($\circ$, $\bullet$) and $IP$
($\square$, $\blacksquare$) for rigid (upper points) and
self-compressed (lower points) spherical cluster of volume 4
$nm^3$  \cite{26} are placed \lq\lq\,between\,\rq\rq\, the slabs
and wires. (c) The size-dependent part of the number of electrons
spilling out from the sample in a contact with reservoirs. (d)
Total kinetic energy $K$ of the electrons (left axis) and
elongation force $F$ (right axis) of the isolated sample.}
\end{figure}
Figure 1(a) shows the density of states, $D(E_{F})$, of an
isolated sample (curve 1) and the one contacted with reservoirs
(curve 2) versus the length of wire. For a better demonstration
the curves 2 are slightly shifted up. The peaks of the $D(E_{F})$
for a wire with square cross-section are more intensive for the
additionally degenerate subband, $k_{n_{y}}=k_{n_{z}}$.

Figure 1(b) displays the behavior of the electron work function
and the ionization potential of the isolated samples of varying
size. The potential difference $\delta \phi $ that arises at the
contact is also indicated. The inequality $W(L)<W_{0}$ is observed
to be obeyed over the whole range of the considered lengths of the
slab and wire. In the shortest wire the five subbands appear and
four of them disappear at elongation. The amplitudes of
oscillations of the Fermi level are of one or two tenths of an eV.
This is substantially less than calculated self-consistently for
the extended thin Al slabs \cite{23} and wires \cite{24,25}. As
can be seen, there are the ranges of widths, where $IP<W_0$ and
$IP>W_0$. The fact that $IP<W_0$ is rather unexpected. Judging
from the empirical fact that the work function $W_0$ of the alkali
metal is approximately equal to one half of the $IP$ of the atom,
one would expect that the $IP$ of a small solid (independently of
the shape of its surface) falls in the range $W_0< IP({\rm
cluster})< IP({\rm atom})$.

The size correction in $W(L)$
competes with the $e^{2}/2C$ term in expression (\ref{IP}) for
the ionization potential. The magnitude of $IP(L)$ depends on the shape of a
sample: $IP(L)>W_{0}$ for $L<43\lambda _{F0}$ and $IP(L)<W_{0}$ for $%
L>43\lambda _{F0}$. Note that the positions of the local minima in the $%
IP(L) $ and $W(L)$ curves correspond to the positions of the peaks
in the density of states of an isolated sample (Fig.1(a)). In the
case $a=L$, we deal with a metal cube whose ionization potential
must have a value similar to that of a sphere. It should be noted
that the ionization potential depends only on the geometry of the
sample and is independent of the direction of electron emission
\cite{30,30a}.

The CPD leads to a noticeable negative shift of the potential well depth
(in utmost point by about 0.5 eV) and this in turn leads (Fig.1(a)) to a
shift in the density of states (curve 2) to a zone of the bigger cross section.

Figure 1(c) shows the size dependence of the number of electrons $\delta N$
that have left the sample. An increase of the cross-section or a decrease in
the wire length leads to the breaks appearing in the monotonic components of
$\delta \phi (L)$ and $\delta N(L)$ which correspond to a spherical sample.
For $a$ of the size close to the atomic diameter the $\delta N$ makes up
20\% of the initial number of electrons. The account for the contact
potential leads to a dependence of $\delta N$ on the hierarchy of energy
levels in an isolated sample. This aligns specific breaks in the $\delta
N(L) $ curve which the positions of the peaks in the density of states $%
D(E_{F})$ of both isolated and contacted samples. The results show that some
part of the electrons is spilled out from the slab (wire) in a contact and
the contact can be thought of as a sort of the ``electron pump'' which pulls
out the electron liquid from or draws it into reservoirs. The dipole layers
formed in a vicinity of both contacts must stimulate additional longitudinal
deformation of the wire and contribute to a change in its shape and in the
electron density. The above results allow to calculate size dependence of
the effective capacitance $C_{c}=e\delta N/\delta \phi $ for a sample in
contact.

The kinetic energy of electrons and the elongation force of the
isolated sample are displayed in Fig.~1(d). Their counterparts for
the sample after junction with reservoirs as well as the potential
$\Omega $ are shown in Fig.~2. Total electron kinetic energy $K(L)$ and
$K_{1}(L)$ of metal cube is similar to that for semi-infinite metal,
$(3/5)NE_{F0}$.
\begin{figure}[!ht]
\centering
\includegraphics [width=0.45\textwidth] {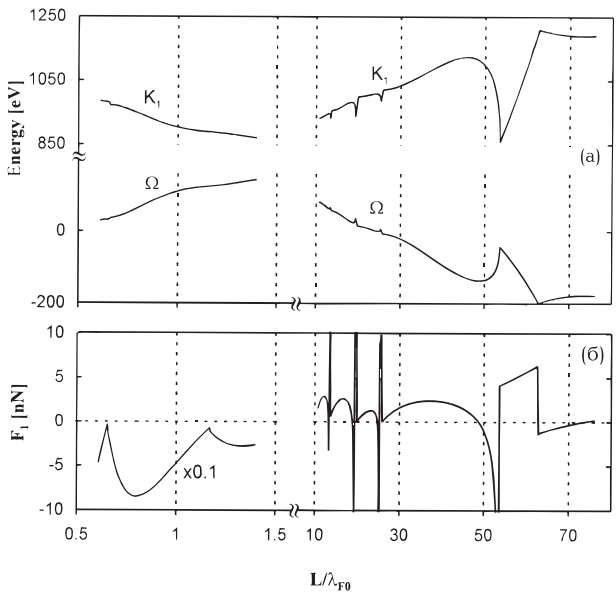}
\caption{The total kinetic energy $K_1$, the grand potential
$\Omega $, and the elastic force $F_1$ of elongated slab and wire
in a contact.}
\end{figure}
The amplitudes of the force oscillations $F(L)$ of the isolated
slab are several times bigger than those measured for the
nanowires \cite{1,9}. It is stipulated by a large value of the
cross-section $a\times a$. Nevertheless, dividing the first
amplitude of the force by the number of atoms in the slab one gets
0.4 nN. This value is smaller than the force acting on a single
atom (1.6 nN) in the point contact experiments \cite{1,9}. That
can be explained by the different dimensionality of a slab and
wire. For the  reduced dimensionality the amplitudes of force
oscillations range from 1 to 3 nN, which is similar to the
measured values. The character of the dependence $F(L)$ changes
significantly at the wire contact because of depletion of the
electron subsystem and due to a shift of the spectrum. It should
be mentioned that in our approach the size dependencies $K(L)$ and
$K_{1}(L)$ are qualitatively different from those obtained for an
infinitely deep potential well \cite{2,3,6}. The second term
appearing in (\ref{PCF2}) only partly smoothes out the
oscillations in $\Omega (L)$. This is in contrast to the
infinitely deep potential well where the oscillations in $\Omega
(L)$ disappear. As a result, the amplitudes of force $F_{1}(L)$
and their shape change significantly. In our more advanced model,
the attempt to take into account the contact potential difference
leads to the appearance of spikes in the elastic force versus
length which result from the fact that for the isolated and
contacted wires the densities of states are not in line. It is
possible that in experiment these peaks could not be recognized in
the background of the thermal fluctuations of the wire-form which
occur during deformation. With the exception of the interval
$54<L/\lambda _{F0} < 62$ (Fig.2(b)) the calculated amplitudes of
force agree with the experimental values. The difference may be
stipulated by not self-consistent determination of the CPD for the
wires of diameter that is comparable with the atomic dimensions
and by the assumption that $L$ varies continuously. Another reason
might be that the elastic force is sensitive to the applied
voltage as suggested in Refs.\ \cite{27,27a}.

The choice of the simplified form of the wire and a non
self-consistent treatment of the size oscillations in the
potential $U$ are the disadvantages of the model. The
free-electron model is not able to describe in detail the effect
of the atomic rearrangements on the magnitude of force in the
atomic contact \cite{9}. On the other hand the obtained
qualitative results are quite general and should not depend on the
symmetry of the problem. It should be also noted that for the
stabilized jellium model the bulk modulus\cite{30a,28}, work
function, surface energy, and surface stress \cite{30,30a} for
gold can be described only when its valency is assumed to be
three. The transition from monovalent to trivalent gold can lead
to increased, by several times, amplitude of oscillations of the
elastic force, in analogy with that observed for the trivalent Al
\cite{13}. These problems can be connected with an unusual size
dependence of $IP(R)$ that is measured for spherical gold
particles: for the very large as well as for the atomic-size
particles $IP>W_{0}$, whereas $IP<W_{0}$ for particles of the
intermediate $R$ \cite{31,32}. In our view, this could be
explained by a reduced valency of some metals which is connected
with the decreasing of the size of a sample (it reminds the
metal-nonmetal transition). A similar transition was observed
\cite{33} for mercury clusters and interpreted in Ref.\cite{34}.
The behavior of the ionization potential measured for aluminium
clusters (compare Fig.28(a) in \cite{35}) also points to this
effect. Therefore, the assumption of monovalency of Au in the
low-dimensional structures is quite reasonable.

Finally we would like point to the connection between our results
and the famous phenomenon of the quantization of shear modulus for
polycrystalline samples that was experimentally observed for 57
elements at the room temperature \cite{37}. Possibly this effect
may be explained by a quantization of the force characteristics of
the 2D electron liquids in a inter-crystallite space of the
polycrystal.

The results of this paper can be summarized as follows. The
evolution of the size dependence of the ionization potential
during transformation of the shape of a sample from a slab toward
a wire is investigated. The size dependence of the CPD is
calculated in a simple manner. It is shown that the infinite
potential well model of a slab \cite{7} overestimates the value of
the CPD by many times. For the first time, the effect of the CPD
on the size oscillations of force was demonstrated. The magnitude
of the calculated CPD is significant for the wires of subatomic
diameters only. An analytical expression for the dependence of the
potential well depth on the Fermi energy for the semi-infinite
metallic system is tested for low-dimensional structures.
[Remarks: Notwithstanding that expressions derived in
\cite{14,15,16,16a} successfully describe character of the
behavior of the electron work function (of positive value) for
semi-infinite metals and ionization potentials for metallic
clusters, in our view, a usage of the classical electrostatics
concept (image force) to define the quantum characteristic - work
function - is not whole appropriate because it cannot
unequivocally determine the sign of the emitted particle. For
example, the positron work function is positive for sodium and
negative for aluminum in spite of identical action of the image
force on this particle \cite{17}.]

\acknowledgments{The work of V.V.P. during his stay in Wroc{\l }aw
was supported by the Mianowski Fund (Poland) and partially by the
Ministry of Education and Science of Ukraine.}



\end{document}